\documentclass[a4paper,12pt]{article}

\usepackage{rotating}
\usepackage{graphicx} 
\usepackage{amsmath}
\usepackage{eurosym}
\usepackage[authoryear]{natbib}
\usepackage[margin=1in]{geometry}
\usepackage[latin1]{inputenc}
\usepackage{appendix}
\usepackage{setspace}

\begin{document}

\title{Global Networks of Trade and Bits\footnotetext{The authors thank participants at the Trade NetWorkShop 3.0 (Trento, December 2010), and at ARS'11 (Naples, June 2011) for insightful comments on an earlier draft of this paper. They blame each other for all other mistakes.}} 
\author{Massimo Riccaboni\thanks{LIME, IMT Institute of Advanced Studies, Lucca. e-mail: mriccaboni@gmail.com} \and Alessandro Rossi\thanks{DISA, University of Trento. e-mail: alessandro.rossi@unitn.it} \and Stefano Schiavo\thanks{DECO, University of Trento and OFCE-DRIC. e-mail: stefano.schiavo@unitn.it}}
\date{\today}
\maketitle

\thispagestyle{empty}

\begin{abstract}
\noindent 
%ALE
Considerable efforts have been made in recent years to produce detailed topologies of the Internet. Although Internet
topology data have been brought to the attention of a wide and somewhat diverse audience of scholars, so far they have
been overlooked by economists. In this paper, we suggest that such data could be effectively treated as a proxy to
characterize the size of the 
%In many cases this has been done by collecting data on network connectivity so to describe the topological properties of the Internet in terms of a graph collected at the Service Provider (ISP) or at the Autonomous System (AS) level.
%The complex and constantly evolving architecture of today's Internet has made it challenging for Internet traffic engineers, network researchers, and other scholars, to provide a truthful map of the network and such datasets have been used for a plurality of research agendas ranging from inferring relationships between ASs to test the effectiveness of alternative communication protocols or new routing algorithms.
``digital economy'' at country level and outsourcing: thus, we analyse the topological structure of the network of trade
in digital services (trade in bits) and compare it with that of the more traditional flow of manufactured goods across
countries. To perform meaningful comparisons across networks with different characteristics, we define a stochastic
benchmark for the number of connections among each country-pair, based on hypergeometric  distribution. Original data
are thus filtered by means of different thresholds, so that we only focus on the strongest links, i.e., statistically
significant links.
%Original data are thus filtered using different thresholds so that we focus our
%attention on the strongest links only, i.e., on links the represent a significant departure from the stochastic
%benchmark.
%Thus, given the total number of connections in the whole network, and the number of incoming/outgoing links for each country (i.e., controlling for network density) we can predict the number of links among each country pair or, alternatively, look at the probability of each cell to contain the observed number of connections. 
We find that trade in bits displays a sparser and less hierarchical network structure, which is more similar to trade in
high-skill manufactured goods than total trade. Lastly, distance plays a more prominent role in shaping the network of
international trade in physical goods than trade in digital services.

\vspace{4pt}

\vspace{4pt}
\end{abstract}
\thispagestyle{empty}

%\newpage
%\thispagestyle{empty}
%\mbox{}

\newpage
\setcounter{page}{1}

\newpage
\section{Introduction}

The impact of the Internet on the global economy has been widely debated. Two main effects have been considered: (a) the ``death of distance'' conjecture \citep{cair97,friedman2005,leamer2007} and (b)
the global digital divide. To test these two claims concerning the impact of distance on information and information-intensive goods, as well as access by less developed countries to the digital web
and the consequences of the digital divide on trade and economic growth, scholars should focus on the co-evolution of trade and computer networks. Such an analysis would also contribute toward
shedding new light on the impact of foreign direct investments (FDIs) and offshoring on trade. An increase in horizontal and vertical FDIs does imply a surge in coordination and information costs.
Thus, we should expect information flows to increase and partially substitute trade flows  \citep{bnve04}.

The standard explanation as to why distance matters for trade is that transport costs increase with distance. Access to information about foreign markets and potential trading partners is also
facilitated by proximity. This implies that major improvements in information and communication technologies (ICTs) should reduce the importance of distance for trade. The ``death of distance''
conjecture has been scrutinized and widely questioned, since there is evidence that the negative impact of distance on trade did not decrease after the Internet revolution
\citep{frwe04,disdier2008,berthelon2008}. Conversely, gravity equation estimates on long-term horizons find an increase in the negative impact of distance on trade \citep{coe2002,brun2005}. In
addition, it has been found that Internet relationships are also constrained by distance \citep{blgo06,mok2007,hortc09}.

Computer scientists have been analysing the structural properties and growth of the world network of computers since the late 1990s \citep{faloutsos99,huberman1999,pastor2004}. More recently, the same
network analysis techniques have been applied to studying the world trade network \citep{serrano2003,garlaschelli2004,garlaschelli2005,fagic08}. Although the two streams of literature adopt the same
methodological approach, the theoretical discussion about the impact of the Internet on world trade has not yet been tested. The main reason why the world networks of computers and trade have never
been jointly analysed has to do with the different levels of aggregation at which the two webs are examined and the lack of methodological tools to compare the dynamics of multiple networks. The
Internet topology has been typically investigated at the level of IP addresses. A complete coverage of the world network is still lacking, and only recently aggregate data at the levels of Autonomous
Systems (AS) and countries have been released. 

%However, despite the growing importance of trade in services, reliable statistics on the international exchange of services are still missing.
Conversely, international statistics on the world trade of goods have long been compiled, and in recent years disaggregated data at the level of single exporting firms have become available. Moreover,
only recently has the literature on complex network analysis started to focus on weighted networks  \citep{newman2004,fagic08,bhattacharya2008} and the co-evolution of multiple interdependent networks 
\citep{buldyrev2010}.

AS topology data seems to represent a very promising data source to measure the economic involvement in the digital economy of the various national economic systems, in a field where traditional data
are difficult to collect or not available at all. For instance, while some statistics on the pervasiveness of ICT and exposure to Internet traffic do exist, figures on the trade of information and
telecommunication services (or even services alone) are lacking  \citep{head2009}.
In this respect, complementing the existing economic figures with Internet topology data inferred at AS level may be
a way to characterize the relative involvement (and its evolution over time) of the various countries in the digital economy, and to map the international trade of digital goods and outsourcing \citep{bhagC04}.\footnote{We refer here to the definition of outsourcing used in \citet[p. 93]{bhagC04}, i.e. ``long-distance purchase of services abroad, principally, but not necessarily, via electronic mediums''.}

It should be noted that the services sector is the largest contributor to a country's economy and is correlated to a country's level of development (ranging from about 50 per cent of GDP, in the case of low-income countries, up to 70 per cent, in the case of high-income ones). 
In addition, over the past two decades the growth of trade in services has grown almost fivefold, while trade in goods has ``only'' done so by a factor of 3.5  \citep{lenn09}. The growing importance of services in national economies and international trade is largely due to an increase in the production of intermediate services (i.e., outsourcing). Besides the economic importance of services activity in general, and services outsourcing in particular, this phenomenon has received a huge amount of attention in the media and in political debates  \citep{haqu07} and the sector has increasingly been included in the framework of multilateral negotiations and regional trade agreements \citep{wto98}.

The economic literature also hosts a lively debate on the role of ICT investments and capital in contributing to the productivity, growth and competitiveness of national systems, in the case of both
developed and developing countries. Some authors have argued that the relative importance of the ICT-producing sector in different countries, and its growth over time, may be a factor contributing to
the differences in growth performance which have recently been observed in several OECD countries  \citep{pilac02}. As such, these are sound arguments advocating the need for policies explicitly
targeting ICT for development. These concerns have also gathered momentum in the agenda of many governments: in response to the recent economic crisis, many countries have implemented policies aimed at containing the cyclical impact of the recession and at reviving the economy. Compared with the past, where interventions gave priority to physical infrastructures, recent policies are distinguished by their greater attention to public intervention in the realms of digital infrastructures and how such a stimulus can affect productivity, innovation, sustainability and the quality of life \citep{anca09}.
%Overall, the first twenty economies (G20) have recently invested 2 trillion dollars, around 100 billion of which in the field of information and communication technologies (Andes and Castro, 2009). 
However, in their present state, many studies on the relationship of ICT investment with growth and productivity pose more questions than answers. For instance, while evidence at sectoral level on
the effects of ICT on productivity and growth for some major countries such as the US and Australia are incontrovertible, for many other OECD countries observed improvements are less straightforward
\citep{vanc08}. 
%Moreover, developing countries struggle with digital divide issues and late pace of ICT adoption. 
Disparities in accessing the Internet are increasingly viewed as an important sources of poverty and inequality, and leveling the playing field in broadband network access is viewed as a key factor in
reducing inequalities  \citep{mehrc04}. One explanation of such different returns on ICT investment may be related to different national capabilities in leveraging those investments in terms of
competitiveness and involvement in international trade. While there is a considerable amount of data on ICT investment and on the international trade of physical goods, almost no data or research
focus on issues such as the flow of digital bits, the international trade in ICT service industries, or how Internet penetration and broadband connectivity affect the competitiveness of economic
systems or the international provision of services  \citep[but see][for notable exceptions]{frwe02,frwe04}.
%The reasons for the lack of research in this field are twofold (IMF,  2003): first, measuring trade in services is harder  than measuring the trade in goods; secondly, there is insufficient  agreement on a detailed taxonomy of ICT- and  Internet-related services (OECD, 2001; IMF, 2003; Lipsey, 2006).  
Starting from such limitations in the literature, in this paper we use the data on Internet interconnectivity as a proxy measure for the International Trade in Service (ITS) and evaluate those data to
measure the involvement of the various countries in the international trade of digital goods and services.

The contribution of the paper is threefold. First, to the best of our knowledge this is the first time that the network of international Internet connectivity has been analysed at country level and
used to proxy the amount of trade in digital services. Second, we provide a detailed comparison between trade in digital and physical goods, and analyse the different role played by distance in
shaping the two networks. The third contribution is methodological and compares networks featuring different characteristics. To perform meaningful comparisons, we propose a stochastic approach based
on hypergeometric distribution.

The paper is organized as follows: in the next section we introduce the various datasets used in this study and, due to their differences (e.g., type of data, methods of data collection), we describe
the methodology used to increase the comparability between them. Section three reviews the major findings arising from network analysis of the Bits and the various Trade datasets. In Section four, we
discuss our results and provide some further conclusive remarks.

\section{Data and Methodology}

\subsection{Internet Data}

In this paper, we make use of data regarding the high-level architecture of the Internet collected within the research activities of the DIMES (Distributed Internet Measurement and Simulations)
project.\footnote{DIMES is a distributed research collaboration project,  which aims  at studying what the Internet looks like and how it evolves over time. The project,  coordinated by a research
team at Tel Aviv University,  has employed over the past few years,  thousands of volunteers in over one hundred countries,  to gather data on Internet topology and has generated one of the most
extensive and reliable datasets freely available for research purposes. See \citet{shsh05} and \texttt{http://www.netdimes.org} for additional information on the project.}

The basic unit of observation is represented by an Autonomous System (AS), which are Internet Service Provider (ISP) or other large subnetworks into which the Internet is divided (these are parts of
the network belonging to a single administrative unit, such as a university, business enterprise, or business division). ASs may be viewed as high-level nodes, so that data exchanges among them are
the edges of a high-level Internet map. Information related to interconnectivity between ASs has remained largely unknown until recently, since ISPs and other business operators are generally not
willing to disclose their connectivity strategies publicly. Nevertheless, composing a truthful map of the Internet is viewed by network researchers as a very important task for routing efficiency,
scalability, security, and many other reasons. As a result, in the past ten years many efforts have been made toward providing reliable measures of AS interconnectivity.

Such attempts of monitoring network packets via traceroute and ping algorithms have reproduced significant parts of the network through random sampling. The DIMES project has implemented these
techniques by means of a distributed multi-point infrastructure (composed of lightweight measurement software hosted by volunteers on computers all over the globe), which has increased both the number
of edges and nodes discovered and improved the representativeness of the data collected, with respect to the whole Internet. Since the beginning of the project (September 1, 2004), the project has
collected 10,833,966,958 independent measurements (records of data exchange between two ASs), occurring between 29,404 ASs distributed in 224 countries and along 204,204 edges, using a total of 24,421
software agents distributed in over 121 countries of the world.

As regards data preparation, the original dataset describes an edge in terms of: (\emph{i}) the city and country of the source and destination nodes; and  (\emph{ii}) the number of distinct couples of
ASs which were observed exchanging traffic between the two cities during a given month. In order to derive the country-country adjacency matrix for the year 2007, we dropped intra-country exchanges and consolidate edges at country and year levels.
%Each node was then associated with an intensity of exchange measure which was built by adding up the figures of the number of distinct couples of IPs for all the corresponding edges at the city and
%country level and for all the monthly samplings. 
The result of this aggregation is a square matrix whose entries show the number of distinct ASs that were observed communicating during the year between any pair of countries.

There are some comments regarding the quality of the dataset which are worth making. As previously mentioned, the data were collected via sampling techniques and do not provide the complete set of interconnections among ASs.
%Reconstructing the whole map of the Internet is considered to be a daunting task, both for the lack of information on the routing policies followed by ASs and for the dynamic and ever changing landscape of interconnections among them.
In spite of this, the DIMES data are based on a large, distributed set of observation points which, with respect to the techniques based on smaller sets of agents or other passive monitoring
techniques, should improve the data in at least three different way: by enlarging the absolute number of discovered nodes and edges; increasing the share of discovered peer-to-peer links (thus
diminishing the bias toward customer-provider links); and minimizing the so-called  ``small observer population'' bias, according to which sampling from a few nodes may result in power-law
distributions of links even in the case of ordinary random graphs.

Regarding the previously defined ``intensity of exchange'' measure, the figures build upon the aggregation of various observations on the interconnections which take place between pairs of ASs in
different countries, in a given period of time. In this sense, although these figures simply signal the existence of distinct connections, we treat them as a rough proxy of the intensity of data
exchange (and by extension, as an even rougher proxy for digital trade). It should be stressed that we are not able to observe the amount of data exchanged (which would be a proper measure of
magnitude of traffic flow). In addition, these figures may be further biased by different practices employed by different ASs as regards the assignment of IP numbers to computers (to name just one
example: an AS might decide to make extensive use of virtual IP addressing, which results in the same machine responding to a plurality of IP numbers, which would increase the intensity figures of our
dataset). Lastly, aggregating monthly values into yearly ones (by simply adding up the figures) may partially counterbalance our inability to distinguish between more and less intense traffic
exchange, since stable, recurring links are weighted more than unstable ones in the resulting measure of exchange intensity.

%Existing datasets on international service trade, which rely on Balance of Payments  transactions between resident and non-resident entities  cover three out of the four modes of international service supply defined in the General Agreement on Trade in Services \citep{oecd08}, namely, cross-border supply (mode 1), consumption abroad (mode 2), and the presence of natural persons (mode 4), leaving out commercial presence (mode 3). If we want to use our data as a proxy for measuring service transactions, we have to keep in mind that they only cover mode 1 and a small fraction of mode 3 transactions, albeit in the context of digital service provision mode 2 and 4 might be very marginal. 
Against this background we should remind that existing international service trade datasets suffer from severe limitations: for instance, the World Bank's World Development Indicators, although
providing the time and country coverage, do not offer bilateral trade data. Of the bilateral datasets, Eurostat only covers 31 countries, and the OECD only four large categories of service, which do
not allow us to focus on the sectoral specificities of the digital economy  \citep{head2009}.

% NB -> info on source-destination, we DO not capture traffic that is just passing through (we neeed to be sure about this as it substantially modifies the interpretation)
% [SONO ABBASTANZA SICURO CHE SI TRATTI DI NODI TERMINALI, sia per destinazione che per origine, E NON DI TRAFFICO DI PASSAGGIO, pero' non ho trovato nessuno che lo dica con chiarezza].

\subsection{Trade Data}
%BACI: world trade database developed by CEPII 
%BACI provides bilateral values (and quantities) of exports at HS 6-digit, for more than 200 countries over the period 1995-2007
%starting from 6-digit data we aggregate to get total trade (Trade\$)
%we also keep information on the number of products traded by each country-pair (TradeN)
%we distinguish between high- and low-skill goods (TradeN-Hi, TradeN-Lo) to investigate whether trade in high-skill goods is more similar to digital trade

We use trade data taken from the BACI dataset maintained by CEPII \citep{gazi10}, which collects information on bilateral trade flows among over 200 countries, at the 6-digit level of the Harmonized
System (HS) classification. In our analysis, we focus on the year 2007, although choosing a different year does not alter the results. Total bilateral trade is obtained by aggregating 6-digit HS flows
for each country pair. Data are expressed in thousands of US dollars, and display a lower cutoff at 1,000 dollars. In addition, to enhance comparability with data on Internet topology, we look at the
number of HS-6 products traded by each country pair, and perform a decomposition of goods according to the skill intensity associated with their production. Starting from  \citet{peneder07} we define
two broad classes of goods, high-skill and low-skill, based on the ISIC sector with which each HS-6 product can be associated.\footnote{\citet{peneder07} provides an international classification of
sectors, ranging between 1 (very high) to 7 (very low), referring to the educational intensity of the sector. In our work, high-skill products are associated with sectors ranging from 1 to 3, i.e., 20
out of the 56 sectors classified by \citet{peneder07}.}

When merging the two datasets, we retain only the countries which are present in both samples, to enhance comparability. We end up with a common set of 189 nodes, populating both trade and Internet
networks.\footnote{The complete list of countries analysed is available in Appendix \ref{sec:countries}.}

\subsection{Methodology}

The way in which the DIMES data was sampled makes a direct comparison between the Bits and Trade networks unreliable. Strong heterogeneity in the number of relationships in the two networks does make
it very difficult to discriminate links which only  reflect the data collection method from those which provide useful information on the properties of the system. To address this issue, we used a
stochastic benchmark for normalization purposes, similar to that recently proposed by  \citet{micciche2011}. This method  is regularly used in genetics to identify statistically significant
relationships  \citep{tavazoie1999}, and a similar approach has been applied to identify statistically significant cliques and clusters in networks  \citep{alba1973,wuchty2006}.

Normalization is the typical strategy used to compare networks: it may be applied to either rows or columns, or to the entire adjacency matrix (for example, rescaling all trade flows as percentages of
the total amount of trade flow in the whole matrix). Normalization may also be applied to both rows and columns, iteratively. For example, if we want an  ``average'' number to put in each cell of the
trade flow matrix, so that both rows and columns add up to 1, we can apply the iterative row and column approach. This is sometimes used when we want to give  ``equal weight" to each node, and to take
into account the structure of both outflows (rows) and inflows (columns). 

Rescaling can be helpful in highlighting the structural features of the data but, obviously, different normalizing approaches highlight different features.

For two countries,  A and  B,  let $N_A$ be the number of goods exported by country A and $N_B$ the number of goods
imported by country B. The total number of traded goods is $N_k$ and the observed number of goods exported from A to B
is $N_{AB}$. Under the null hypothesis of random co-occurrence, that is to say, customers in country B are indifferent to
the nationality of the exporter, the probability of observing $X$ goods traded is given by the hypergeometric distribution \citep{Feller}
\begin{equation}
H(X|N_k,N_A,N_B)=\frac{{N_A \choose X}{N_k-N_A \choose N_B-X}}{{N_k \choose N_B}}.
\label{hyper}
\end{equation}

We can associate a $p$-value with the observed $N_{AB}$ as
\begin{equation}
p(N_{AB}) =1- \sum_{X=0}^{N_{AB}-1} H(X|N_k,N_A,N_B).
\end{equation}
Note that the described null hypothesis directly takes into account the heterogeneity
of countries with respect to the number of goods traded (row and column totals). For each pair of countries, we separately evaluate the $p$-value for each trade relationship and
%we count the number of country pairs in which the $p$-value is smaller than a selected statistical threshold. We accordingly set a link between A and B if the number of goods traded is not vanishing and we use it as the weight of the link. We 
then use different cutoffs to select only those links which represent a significant departure from the hypergeometric benchmark ($p<.05, p<10^{-12}$). The resulting matrices are then dichotomized. The
hypergeometric multi-urn benchmark is equivalent to the Monte Carlo degree-preserving network rewiring procedure \citep{maslov2002}.\footnote{Alternative normalization procedure has been applied to
trade networks \citep{serrano2007}. In particular, departures from the Gravity benchmark were examined \citep{krempel2003}.} Thanks to our stochastic approach we can treat a weighted network
as a random graph in which any link has a probability of occurrence.

%XXX huh? XXX

\section{Results} We first describe here the results on the topological properties of the Internet and Trade networks  (Section \ref{sec:topo}) and then analyse the role of distance in shaping bilateral links  (Section \ref{sec:dist}).

\subsection{Topological properties}
\label{sec:topo}

\paragraph{Density}

As the data on Internet connectivity were collected via sampling and do not represent the universe of AS links, we expect the Bits network to be less dense than the trade one. This is indeed one of
the issues which led us to use a stochastic approach before running comparative analyses. Table \ref{tab:density} lists the relevant information on the density of the various networks, plus the share
of bilateral (i.e., reciprocal) connections (in brackets).

Density is indeed much higher for Trade than for Bits in the original data (64 Vs 13 percent). Although reduced, this difference remains very large, even for statistically significant ties based on
hypergeometric distribution (3.5 Vs 14 percent, and 2.1 Vs 8.6 percent respectively).

When considering the distinction between high- and low-skill products, we do observe that the density of the first network is lower, in line with our hypothesis that products associated with higher
skill requirements are more similar to trade in digital services. It is interesting to note that, when focusing on the most demanding threshold (i.e., only those links which take on values with a
probability of below  $10^{-12}$), the difference between high- and low-skill products becomes particularly important, as density for the former is half that of the latter, which remains very similar
to the density of the whole trade network. However, the digital network always remains less dense.

\begin{table}[htbp]
  \centering
  \caption{Network density and share of bilateral links}
  \label{tab:density}
  \footnotesize
    \begin{tabular}{lrrrr}
\hline
          & Bits  & Trade & Trade-Hi & Trade-Lo \\
\hline
real  		& 13.382 	& 64.055 	& 49.983 	& 61.511 \\
      		& (69.480) 	& (86.760) 	& (79.890) 	& (85.210) \\
cut 0.05 	& 3.462 	& 13.999 	& 11.393 	& 13.610 \\
     		& (43.250) 	& (61.080) 	& (58.790) 	& (60.170) \\
cut $10^{-12}$	& 2.142 	& 8.646 	& 4.889 	& 8.190 \\
     		& (36.790) 	& (60.220) 	& (51.240) 	& (59.450) \\
\hline
\multicolumn{5}{l}{Data expressed in percentage points} \\
\multicolumn{5}{l}{Figures in brackets: share of bilateral links.} \\
\end{tabular}
\end{table}

Bilateral links represent the majority of connections, and this is particularly true for trade in goods, where less than 15 percent of links are not reciprocated. This share doubles for Internet
connections, and once again the difference remains sizable after hypergeometric filtering.

This finding ---that  the Bits network is less reciprocal than the Trade one--- is not surprising, when we consider that there are global content providers located in a few countries (e.g., the USA)
which distribute contents worldwide, generating large asymmetries in the balance of the flow of bits with respect to countries which are more marginal in production and distribution of digital
contents. Thus, in the case of such disparities, sampling measurements may be unable to report any activity in one direction of flow between two countries.\footnote{In addition, in the case of the
Bits network, there is nothing similar to the trade balance, so disparities in fluxes in principle do not create any imbalance for a country.}

Network density figures suggest that many more country pairs are involved in world trade than in Internet connectivity. Although it is probably true that more countries participate in trade in goods,
not least because this is a much older activity, a lower density for the AS network also suggests that the Internet is more polarized, with strong connections among a smaller set of actors. To further
examine this conjecture, we now turn to other network statistics.

\paragraph{Node Degree}

The average total degree is much higher for the Trade network than for Bits, as one would expect from the higher density of the former. Again, hypergeometric filtering reduces this difference, but
does not substantially alter the picture, and we continue to observe that, while the network of low-skill products is very close to the whole trade network, high-skill goods display behavior that is
more similar to that of the Internet. This similarity is reinforced by hypergeometric filtering.

On average, each country has approximately 136 trade partners, but it connects to ASs located in just over 30 foreign countries, as Table \ref{tab:degree}shows. When we focus only on the strongest
links, we find an average number of 22.72 trade partners compared with only 6.57 connections with foreign ASs. Remarkably, as mentioned above, countries involved in trade of high-skill products on
average only exchange them with 13.67 other countries.

\begin{table}[htbp]
  \centering
  \caption{Summary Information on Degree Connectivity}
    \begin{tabular}{lrrrr}
\hline
          		& \multicolumn{ 4}{c}{Average Total Degree} \\
\cline{2-5}
          		& Bits  	& Trade   & Trade-Hi& Trade-Lo \\
\hline
real  			& 32.836 	& 136.370 & 112.868 & 132.741 \\
cut 0.05 		& 10.201 	& 36.561  & 30.243 	& 35.778 \\
cut $10^{-12}$ 	& 6.571 	& 22.720  & 13.672 	& 21.640 \\
          		&       &       &       &  \\
          		& \multicolumn{ 4}{c}{Average In/Out Degree} \\
\cline{2-5}
          		& Bits   & Trade 	& Trade-Hi	& Trade-Lo \\
\hline
real  			& 25.159 & 120.423  & 93.968 	& 115.640 \\
cut 0.05 		& 6.508  & 26.318 	& 21.418 	& 25.587 \\
cut $10^{-12}$ 	& 4.027  & 16.254 	& 9.191 	& 15.397 \\
          		&        &       	&       	&  \\
				& \multicolumn{ 4}{c}{Max/Min Total Degree} \\
\cline{2-5}
       	  		& Bits   & Trade 	& Trade-Hi 	& Trade-Lo \\
\hline
real  			& 171 / 1& 188 / 28 & 188 / 14 	& 188 / 26 \\
cut 0.05 		& 111 / 0& 125 /  8 & 116 /  5 	& 127 / 7 \\
cut $10^{-12}$ 	&  80 / 0& 106 /  2 &  54 /  1 	& 103 / 2 \\
\hline
    \end{tabular}
  \label{tab:degree}
\end{table}

Differentiating between in- and out-degree (imports and exports) does not substantially alter the picture. On average, the number of in- and out-partners coincides because, for the world as a whole,
every exporting relationship needs to have an importing counterpart.

The bottom panel of Table  \ref{tab:degree} displays the maximum and minimum numbers of partners in each of the networks. Here the difference is less severe, as both Trade and Bits networks have
``extreme'' cases of very well-connected and very poorly connected players. However, in the Bits network, none of the countries is fully connected (whereas this does happen in the case of goods). The
sharper differences concern the minimum degree, as the least connected country in the world trade web still exchanges goods with some 28 partners. The effect of thresholds becomes very evident here:
by pruning not statistically significant  links, hypergeometric filtering makes the minimum total degree very similar across the different networks.

\paragraph{Degree Distribution}

\begin{figure}[htbf]
\centering
\includegraphics[width=6in]{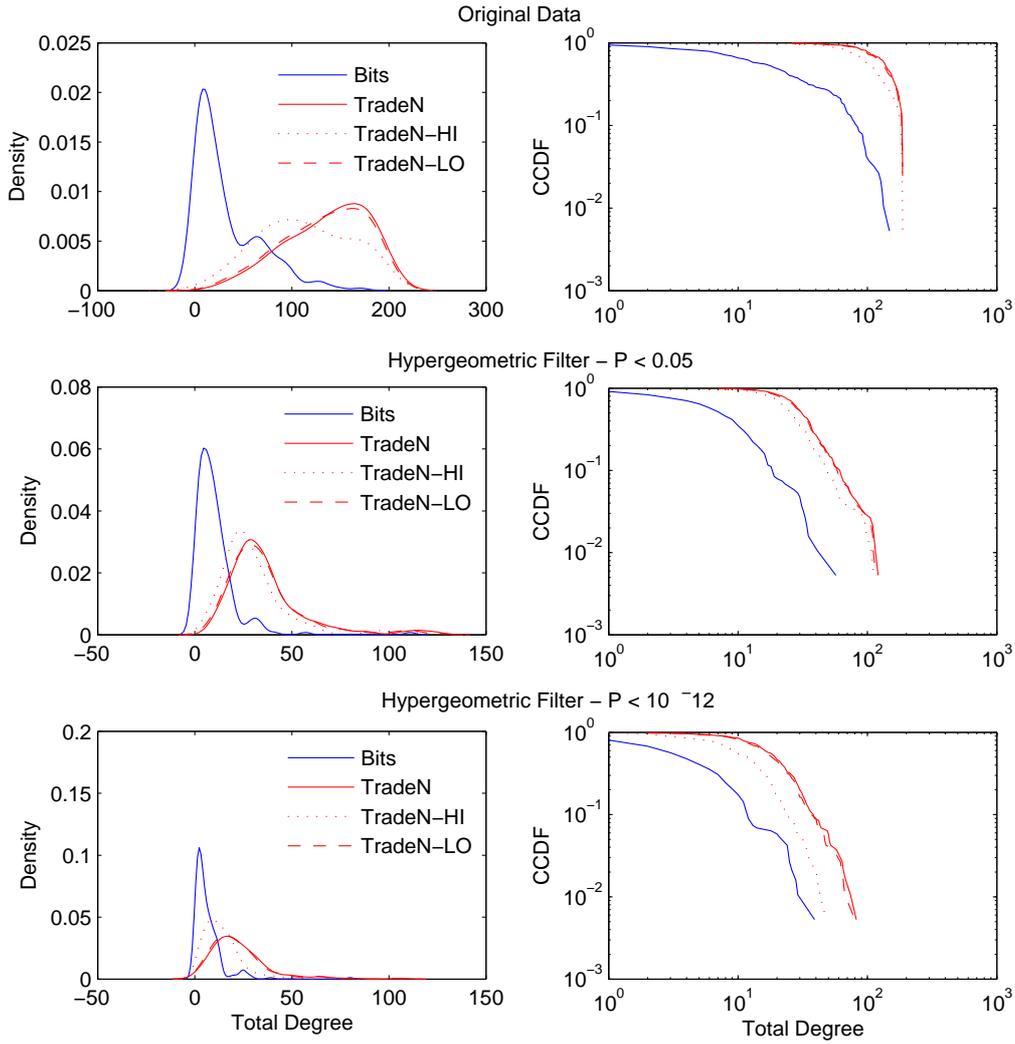}
\caption{Degree distribution. Left panels: kernel density; right panels: counter cumulative distribution (log-log scale). Top row: original data; middle row: hypergeometric filtering with $P<0.05$; bottom row: hypergeometric filtering with $P<10^{-12}$}
\label{fig:degdistr}
\end{figure}

To clarify the different behavior of node degree across the various networks, Figure \ref{fig:degdistr} shows the distribution of node total degrees for the four networks. It shows both the estimated
kernel density (left), and the counter-cumulative distribution function (CCDF) in double-log scale (right).

The Bits network is characterized by a much more skewed distribution at all levels of observation, whereas trade in low-skill goods is almost indistinguishable from the whole trade network, and the
network of high-tech goods lies somewhere in between. A comparison of the top row (original data) with the distributions obtained after applying the hypergeometric filtering, highlights the importance
of the latter in comparing networks with different features adequately. Indeed, the shape of the degree distribution for the various trade networks changes substantially once we apply the cutoffs, and
the difference with respect to the Bits network is reduced.

None of the CCDF approaches the straight line which represents a power-law distribution, although the right tail of the CCDF for Bits is characterized by portions which resemble a power-law.

\paragraph{Assortativity}

So far we have only looked at single-node characteristics, from which we cannot infer much information about the structure of the links. We turn now to the second-order features of the graphs, by
looking at the assortative structure of the networks.

Table \ref{tab:pears} lists the simplest characterization of assortativity, i.e., Pearson's degree correlation. We see that all the networks display a disassortative structure, in which nodes with low
degree tend to link with high-degree partners.\footnote{This is consistent with previous results. See for instance \citet{mahac06} on the Internet topology, and \citet{bhattacharya2008,fagic08,fagiC09pre} on trade.} We can also appreciate the importance of the filtering procedure, as in the original data the Bits network has the larger correlation (in absolute values), while the ranking is reversed once we focus only on the strongest links.

This result implies that, in the Bits network, many of the core-periphery links (i.e., those between high- and low-degree nodes) tend to be weak and do not stand up to  hypergeometric filtering.

\begin{table}[htbp]
\footnotesize
\centering
\caption{Pearson's degree correlation}
\label{tab:pears}
\begin{tabular}{lrrrr}
\hline
      			& Bits  & Trade & Trade-Hi & Trade-Lo \\
\hline
real  			& -0.407 & -0.311 & -0.355 & -0.312 \\
cut 0.05 		& -0.177 & -0.270 & -0.285 & -0.273 \\
cut $10^{-12}$ 	& -0.165 & -0.289 & -0.270 & -0.288 \\
\hline
\end{tabular}
\end{table}

\paragraph{Average K-neighbors degree}

A less synthetic way of looking at assortativity is K- neighbor degree (Knn), i.e., the average degree of partners of nodes with a given degree. Figure 2 plots Knn for the various networks. Although
the three trade networks are very similar, the top-left panel shows that, when looking at the strongest link only (using the most stringent threshold), the Bits network has a flatter Knn profile,
signalling that the average degree of each country's partners is less dependent on its degree than in the case of trade.

\begin{figure}[htbf]
\centering
\includegraphics[width=6in]{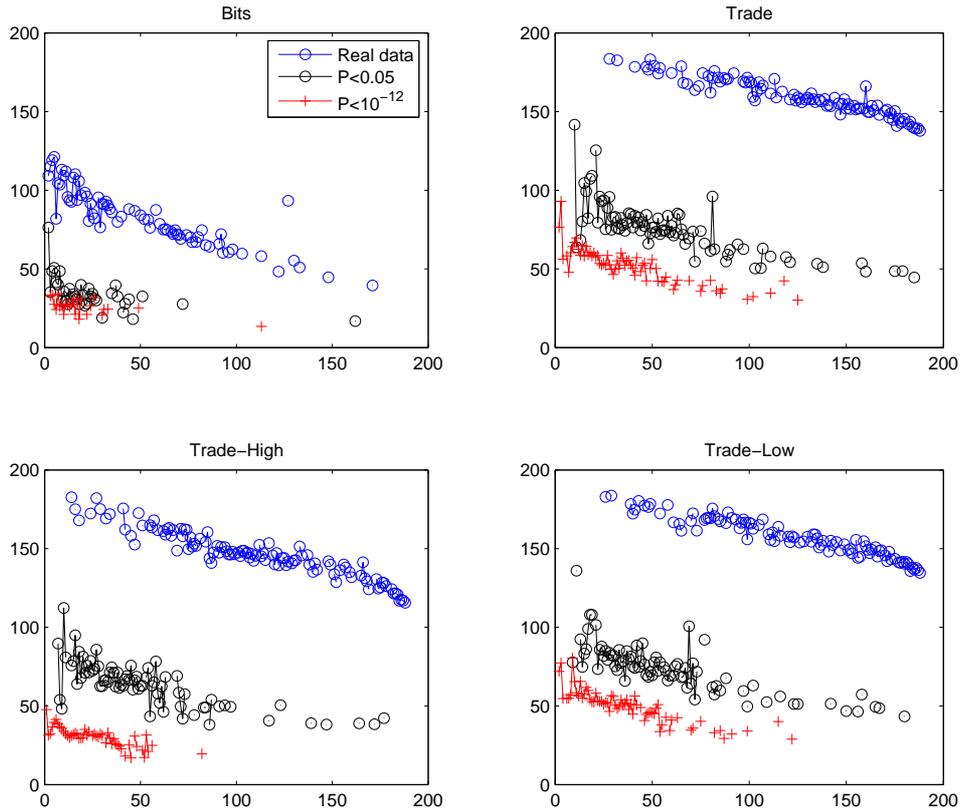}
\caption{K-neighbors degree (Knn): the average degree of nodes with a given degree. Four panels show Knn for various networks: from left to right and from top to bottom Bits, Trade, Trade in high-skills goods and Trade in low-skills goods.}
\label{fig:knn}
\end{figure}

\paragraph{Degree -- Average Nearest Neighbor Degree Correlation}

This feature also emerges from another measure of assortativity, i.e., the correlation between node degree and the average degree of its neighbors (ANND). Table 4 distinguishes between in- and out-degrees and lists correlations between the out-degree of a node and the average in-degree of its partners (top panel), as well as that between the in-degree of a country and the average out-degree of its partners (bottom panel). 
These measures can define whether large-scale exporters tend to ship goods to countries which import from many sources and, vice versa, whether countries which import from many suppliers do so prevalently from countries which serve many customers.

\begin{table}
\centering
\footnotesize
\caption{Correlation Node Degree -- ANND}
\label{tab:annd}
\begin{tabular}{lrrrr}
\hline
          		& \multicolumn{ 4}{c}{Out Degree -- $ANND^{out}_{in}$} \\
\cline{2-5}
          		& Bits  	& Trade   & Trade-Hi& Trade-Lo \\
\hline
real  			& -0.271 & -0.973 & -0.961 & -0.968 \\
cut 0.05 		& -0.141 & -0.647 & -0.651 & -0.624 \\
cut $10^{-12}$ 	& ~0.018 & -0.325 & -0.067 & -0.276 \\
      &       &       &       &  \\
          		& \multicolumn{ 4}{c}{In Degree -- $ANND^{in}_{out}$} \\
\cline{2-5}
          		& Bits  	& Trade   & Trade-Hi& Trade-Lo \\
\hline
real  			& -0.629 & -0.982 & -0.972 & -0.978 \\
cut 0.05 		& -0.235 & -0.684 & -0.620 & -0.705 \\
cut $10^{-12}$ 	& -0.075 & -0.704 & -0.522 & -0.713 \\
\hline
\multicolumn{5}{l}{All correlations are significant at 1\%.} \\
\end{tabular}
\end{table}

We find that the Trade networks feature a much stronger negative correlation between node degree and ANND, but there is also a substantial difference between the top and bottom panels of Table
\ref{tab:annd}. In the former case, when looking at the relation between the number of export markets served and the average number of suppliers they have, the negative correlation for high-skill
goods is very close to zero, whereas  the figure for the Bits network becomes slightly positive. Hence, for both digital services and trade in high-skill goods, we do not find a strong link between
the number of export partners and the number of import relationships which they maintain.

The same is not completely true when we reverse the direction of the edges and look at the correlation between in-degree and $ANND^{in}_{out}$ (bottom panel of  Table \ref{tab:annd}). In this case,
Bits maintain a weak (negative) correlation, whereas Trade-Hi displays a quite strong one. In addition, the correlations are always stronger in the bottom panel than in the top one.

%\paragraph{Joint Degree Distributions}

%\subsubsection*{Cut at 5\%}
%\includegraphics [width=5in]{Ste_June2011_P_02.jpg}
%\includegraphics [width=5in]{Ste_June2011_P_03.jpg}
%\includegraphics [width=5in]{Ste_June2011_P_04.jpg}
%\includegraphics [width=5in]{Ste_June2011_P_05.jpg}
%\includegraphics [width=5in]{Ste_June2011_P_06.jpg}
%\includegraphics [width=5in]{Ste_June2011_P_07.jpg}
%\includegraphics [width=5in]{Ste_June2011_P_08.jpg}

%\subsubsection*{Cut at 0\%}
%\includegraphics [width=5in]{Ste_June2011_P_09.jpg}
%\includegraphics [width=5in]{Ste_June2011_P_10.jpg}
%\includegraphics [width=5in]{Ste_June2011_P_11.jpg}
%\includegraphics [width=5in]{Ste_June2011_P_12.jpg}
%\includegraphics [width=5in]{Ste_June2011_P_13.jpg}
%\includegraphics [width=5in]{Ste_June2011_P_14.jpg}
%\includegraphics [width=5in]{Ste_June2011_P_15.jpg}

\paragraph{Clustering}

Assortativity provides us with some information about the structure of the network, by telling us whether nodes with high degree tend to connect to partners with a similar number of connections. To
understand how these partners interact among themselves we can look at clustering, which tells us how likely it is for someone's partners to be also linked.

On average, the clustering coefficient is quite high in all networks built from the original data, but its value decreases markedly once we control for the hypergeometric benchmark, meaning that many
of the triangles observed in the network show weak ties which do not stand up to the hypergeometric test (see Table  \ref{tab:cc}). This appears to be particularly true for the Bits network.
Interestingly, here there is not much difference across types of goods, since high- and low-skill goods differ very little, whereas the Bits network is much less clustered.

\begin{table}[htbp]
\footnotesize
\centering
\caption{Average clustering coefficient}
\label{tab:cc}
\begin{tabular}{lrrrr}
\hline
      			& Bits  & Trade & Trade-Hi & Trade-Lo \\
\hline
real  			& 0.726 & 0.835 & 0.798 & 0.827 \\
cut 0.05 		& 0.288 & 0.448 & 0.440 & 0.439 \\
cut $10^{-12}$ 	& 0.234 & 0.465 & 0.444 & 0.460 \\
\hline
\end{tabular}
\end{table}

The correlation between node degree and clustering is again negative and significant, much stronger for the three Trade networks than for the Bits case. When looking at the results pertaining to the
stricter threshold, the correlation for Bits is close to zero. Once again, this confirms that trade in digital services, as opposed to trade in physical goods, is characterized by a flatter, less
hierarchical structure.

\begin{table}[htbp]
\footnotesize
\centering
\caption{Correlation Node Degree -- Clustering}
\label{tab:cc2}
\begin{tabular}{lrrrr}
\hline
      			& Bits  & Trade & Trade-Hi & Trade-Lo \\
\hline
real  			&-0.349 & -0.930 & -0.957 & -0.932 \\
cut 0.05 		&-0.180 & -0.701 & -0.639 & -0.663 \\
cut $10^{-12}$	&-0.024 & -0.562 & -0.391 & -0.531 \\
\hline
\end{tabular}
\end{table}

\paragraph{RICH CLUB}

A simple way of characterizing the hierarchical structure of a network is provided by the rich-club coefficient  \citep{zumo04,colic06}.

\begin{figure}[htbf]
\centering
\includegraphics[width=6in]{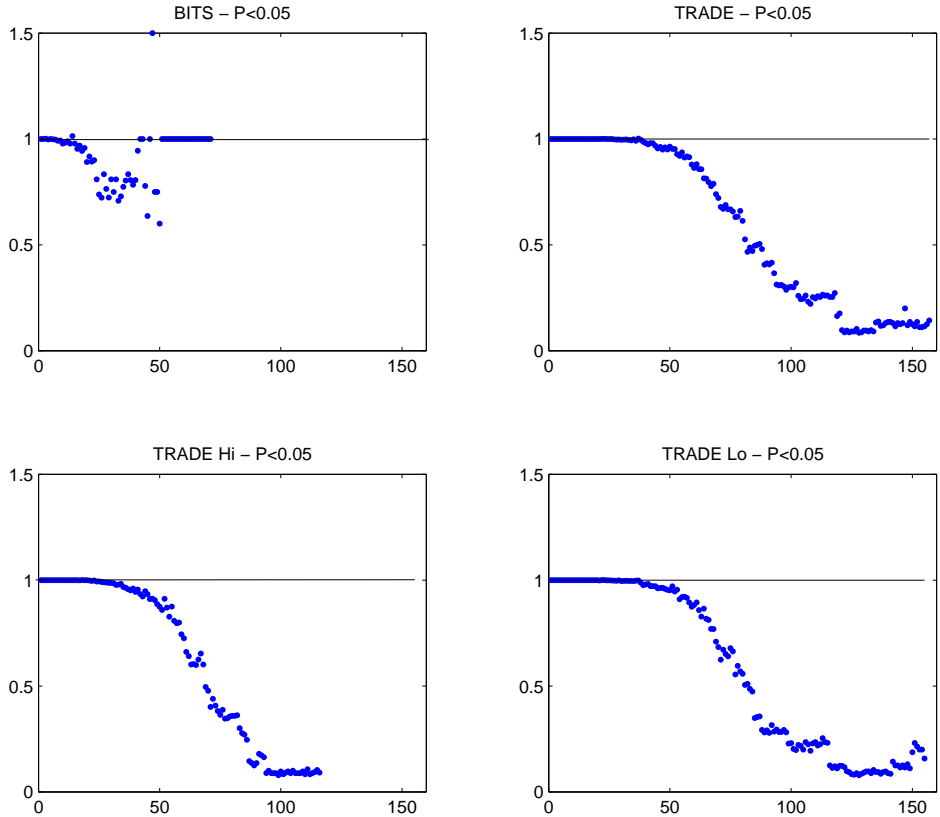}
\caption{Rich-club coefficient on the networks obtained by applying the 5\% cutoff.}
\label{fig:rcc}
\end{figure}

Figure \ref{fig:rcc} shows the results for the rich-club coefficient, computed following  \citet{colic06} for the various networks obtained using the 5 percent hypergeometric cutoff.\footnote{Results
for the original data and the other cutoff value are not shown but are available upon request.} The original rich-club coefficient is compared with that obtained for a random network with a similar
degree sequence, and meaningful rich-club ordering is observed whenever the normalized coefficient is greater than 1. In all cases, we find a normalized coefficient which is smaller than what we would
expect from a random network, suggesting that high-degree nodes (hubs) provide connectivity to regional/local hubs, which in turn are not tightly interconnected. Remarkably, this behavior is much more
pronounced in Trade than in Bits, suggesting that stronger links (compared with our null model based on hypergeometric distribution) occurs mainly between the core and the periphery in trade, whereas
a larger number of links are found among hubs in the case of the Internet.

We conjecture that this behavior depends on the different role played by distance in hindering link formation in the two types of networks (physical versus digital trade). We now explore this issue.

\subsection{The Role of Distance}
\label{sec:dist}

We can examine the role of distance in shaping international trade relationships in several ways. The simplest is to compare average distance across connected pairs in the various domains, as shown in
Table \ref{tab:avgdist}.\footnote{Distance is defined as the geodesic distance between the largest cities/agglomerations (in terms of population) of two countries. We use data provided by CEPII and available at
http://www.cepii.fr/anglaisgraph/bdd/distances.htm .} We see that, in the original data, trade partners are on average further away from each other than country-pairs connected by the exchange of Bits. 
However, once we control for the null model and focus only on the strongest links, the hierarchy is reversed: this is more apparent when moving from the 0.05 threshold to the more stringent cutoff of  $10^{-12}$. The last row of Table \ref{tab:avgdist} does show that average distance is smaller in the three trade networks than in the case of Bits. Hence, stronger links appear to occur at shorter distances, as expected, but this phenomenon is especially marked in the case of trade in physical goods. Note also that, in this case, the similarity between Bits and trade in high-skill goods
which we found previously, does not hold, as Trade-Hi undergoes a very sharp drop in average distance between connected pairs from the complete dataset to the most stringent threshold.

\begin{table}[htbp]
  \centering
  \caption{Average distance between connected country-pairs}
    \begin{tabular}{lrrrrrr}
\hline
          & Bits  & Only bits & Trade & Only trade & Trade-Hi & Trade-Lo \\
\hline
real  			& 6904  &  8113 & 7430 & 7570 & 7145  & 7366 \\
cut 0.05 		& 4594  & 6900 &  4460  & 4662 & 4132  & 4472 \\
cut $10^{-12}$ 	& 4128  & 6288 & 3671  & 3794 & 2948  & 3632 \\
\hline
\multicolumn{7}{l}{Figures in kilometers. Average distance between all countries: Km 8311.} \\
    \end{tabular}
  \label{tab:avgdist}
\end{table}

\begin{figure}[h!]
\centering
\includegraphics[width=8in,angle=90]{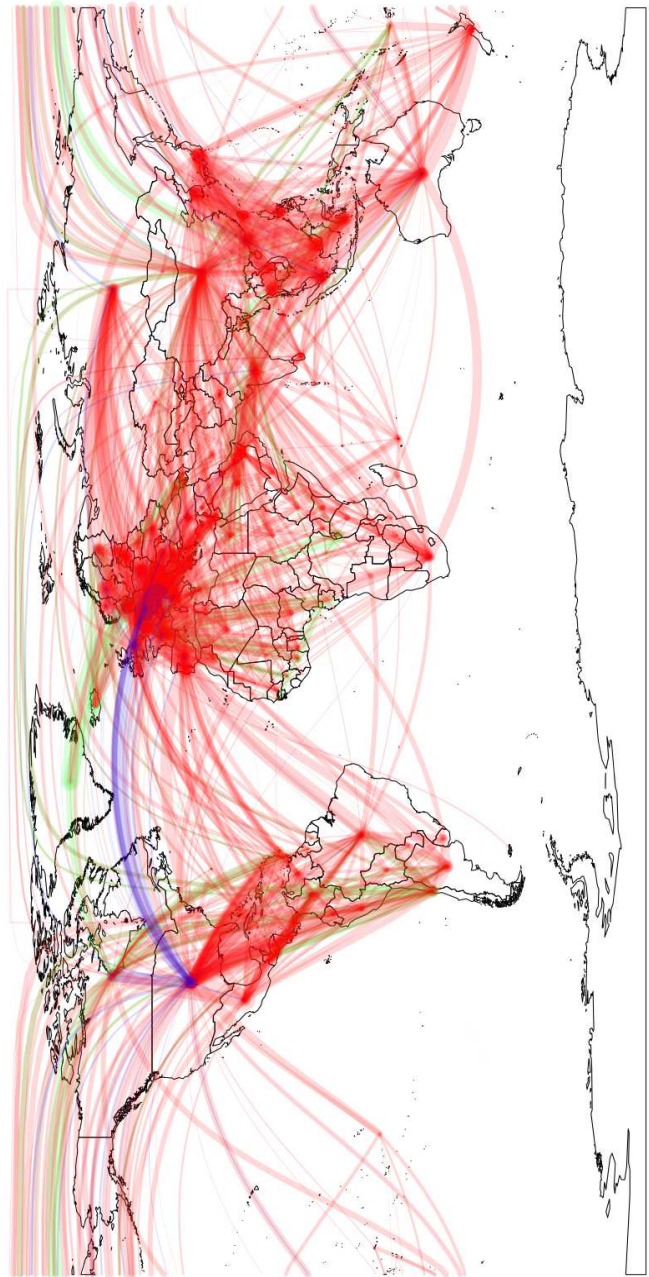} 
\caption{Global computer and trade networks, statistically significant linkages, cut $10^{-12}$. Blue edges: Bits-only connections; red edges: Trade-only connections, green edges: Bits and Trade connections.}
%Blue edges are links which are significant only in the Bits network, red ones represent Trade-only connections, whereas green edges are persisting links in both networks.} 
\label{fig:map2}
\end{figure}

Figure \ref{fig:map2} provides graphic evidence of the different role of distance in the Trade and Bits networks (with the most stringent hypergeometric threshold, blue edges are links which resist
filtering only in the Bits network but are canceled in the Trade network and vice versa for red edges; green edges are persisting links in both networks). The figure  shows how the strongest links
have a geographical (regional) component for trade in physical goods.

To test the fact that average distances actually differ, we revert to a Wilcoxon signed-rank test  \citep{wilc45}, as we cannot assume the population to be normally distributed and use the standard
$t-$test. We find that the null hypothesis of equality of means is rejected for the original data, but cannot be rejected in the case of the strongest links only. This is consistently with the
observation that the average distance for trade links falls down markedly when we trim the trivial links.

\begin{figure}[htbf]
\centering
\includegraphics[width=6.5in]{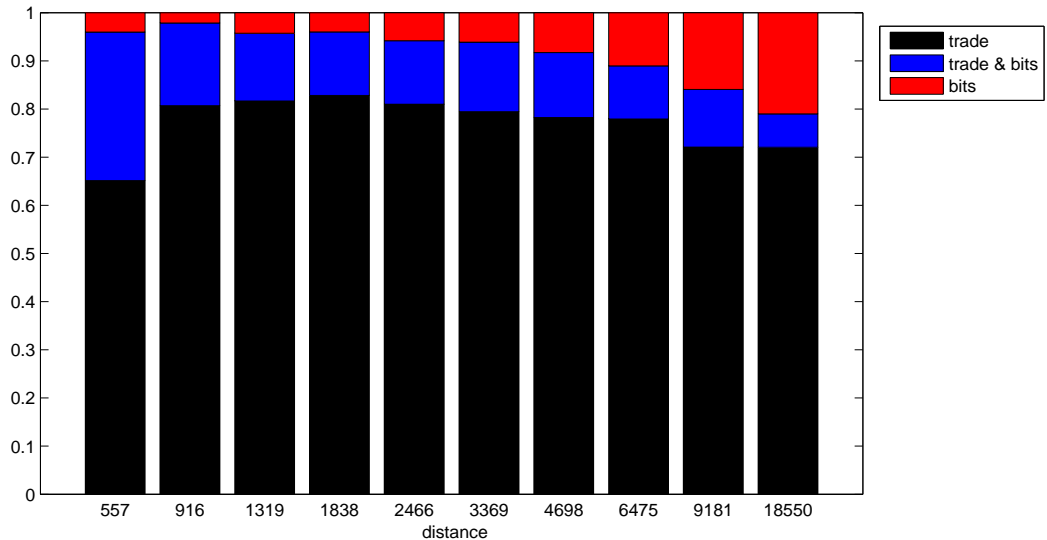}
\caption{Distribution of statistically significant links by type and distance, cut  $10^{-12}$.}
\label{fig:dis}
\end{figure}

Figure \ref{fig:dis} shows that, as distance increases, so does the share of statistically significant links in the Bits networks, while the relative importance of Trade-only connections falls.
If we assume that digital connections lower the transaction costs usually associated with distance, then increased connectivity can tilt the proximity-concentration trade-off that needs to be evaluated when comparing the relative merits of export Vs FDI \citep{brai97,mave00,helpC04}.
In this respect, higher digital connectivity can emerge both as a precondition (lower transaction costs spur multinational activities) and as a result (connections with the headquarter) of higher bilateral FDI flows.
These, in turn, are associated with higher trade if vertical FDI dominates (as intermediate inputs are shipped back and forth), with lower trade if FDI are horizontal, i.e. if they represent an alternative way to serve foreign markets.
Therefore, from a theoretical point of view, the relation between digital connectivity and trade can be either positive or negative.
The fact that in the analysis summarized in Figure \ref{fig:dis} we find a decline in the relative importance of Trade-only and Trade-Bits connections as distance grows (together with an increase in the number of significant links for the Bits network) suggests that at longer distances horizontal FDI are facilitated by easier communications brought about by the ICT revolution, so that traditional export activities tend to be substituted by new forms of serving foreign markets.

\begin{figure}[thbf]
\centering
\begin{tabular}{ccccc}
\multicolumn{5}{c}{}\\
\includegraphics[width=1.5in,height=1.5in,angle=0]{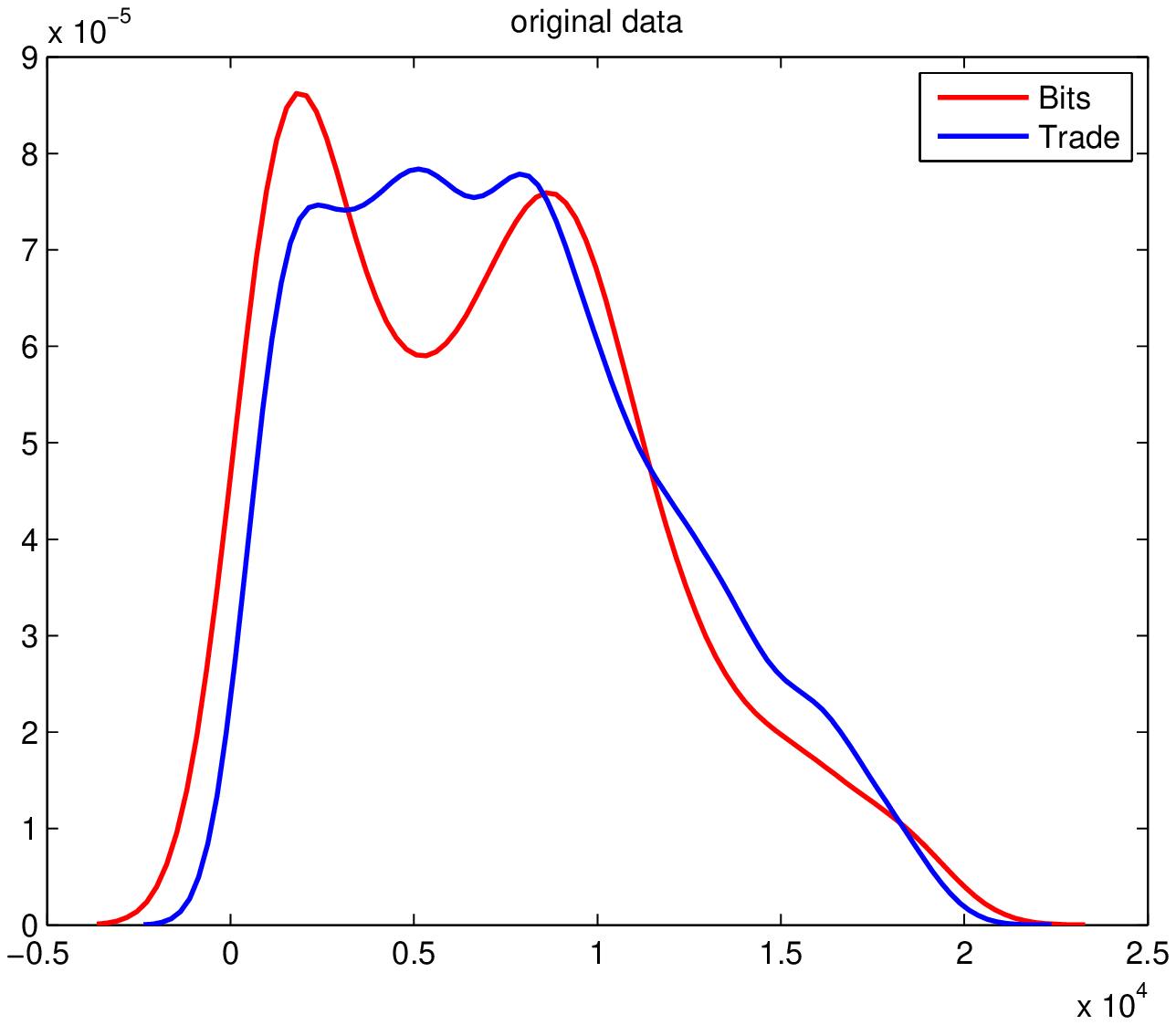} & &
\includegraphics[width=1.5in,height=1.5in,angle=0]{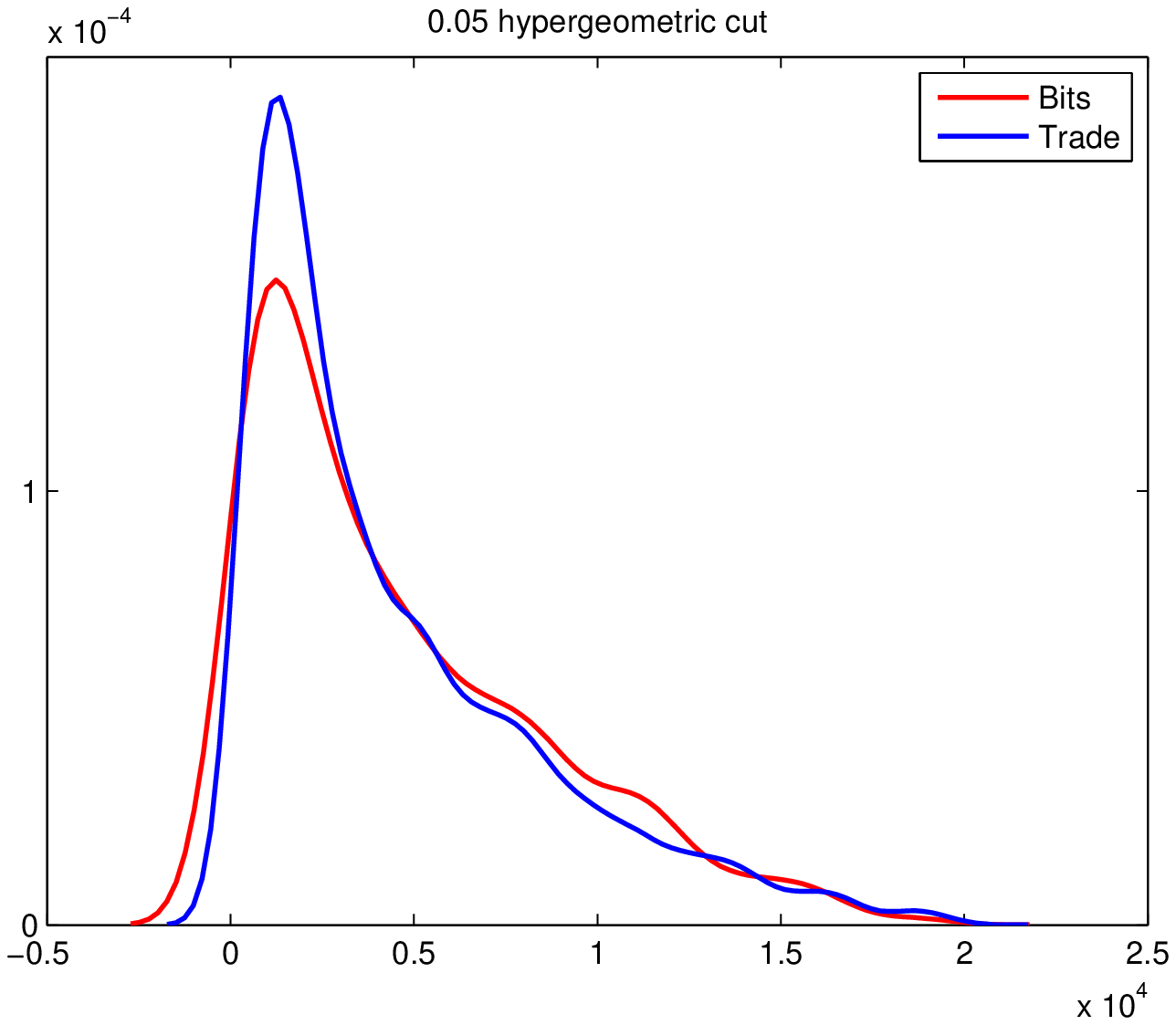} & &
\includegraphics[width=1.5in,height=1.5in,angle=0]{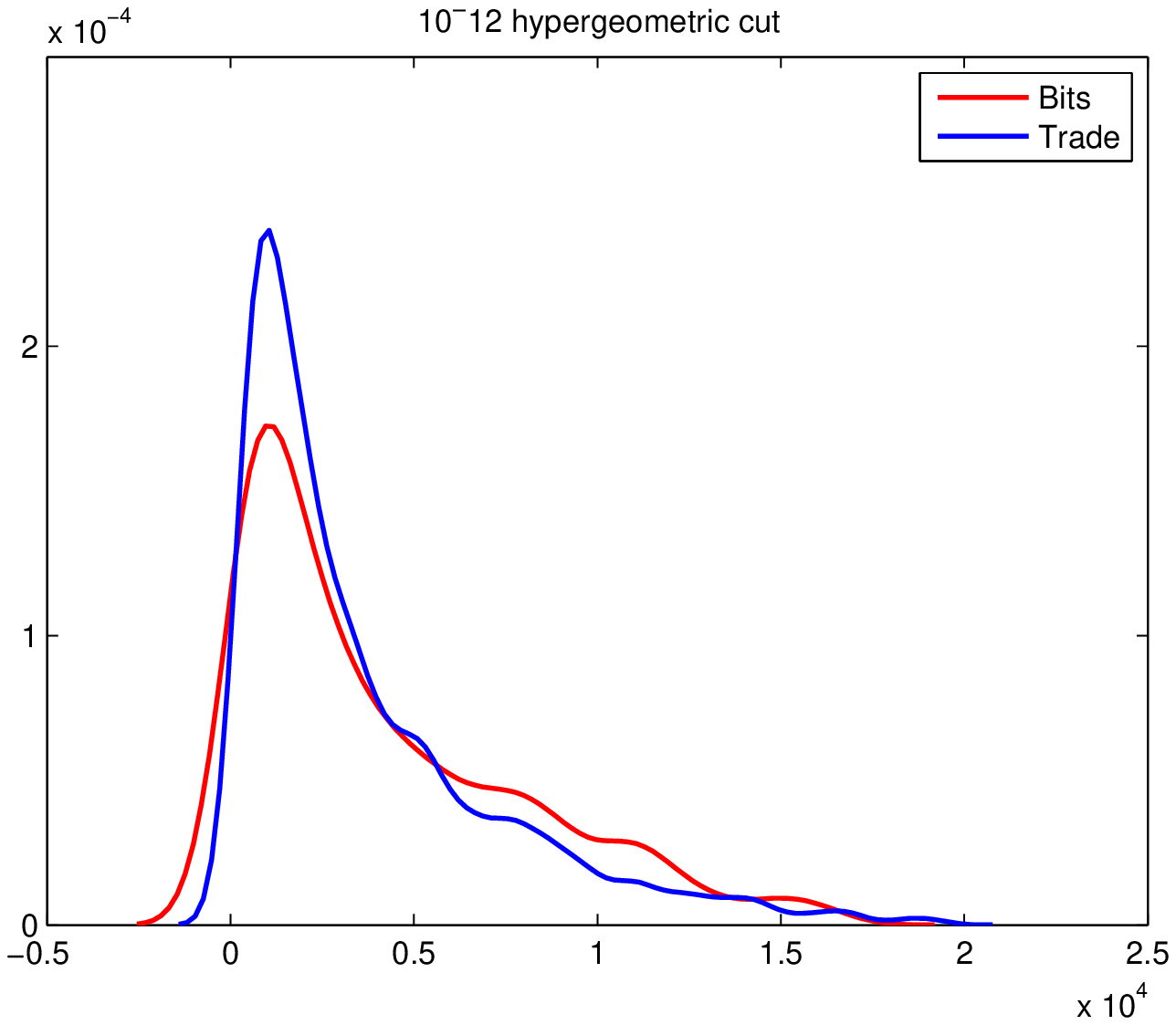} \\
\multicolumn{5}{c}{}\\
\includegraphics[width=1.5in,height=1.5in,angle=0]{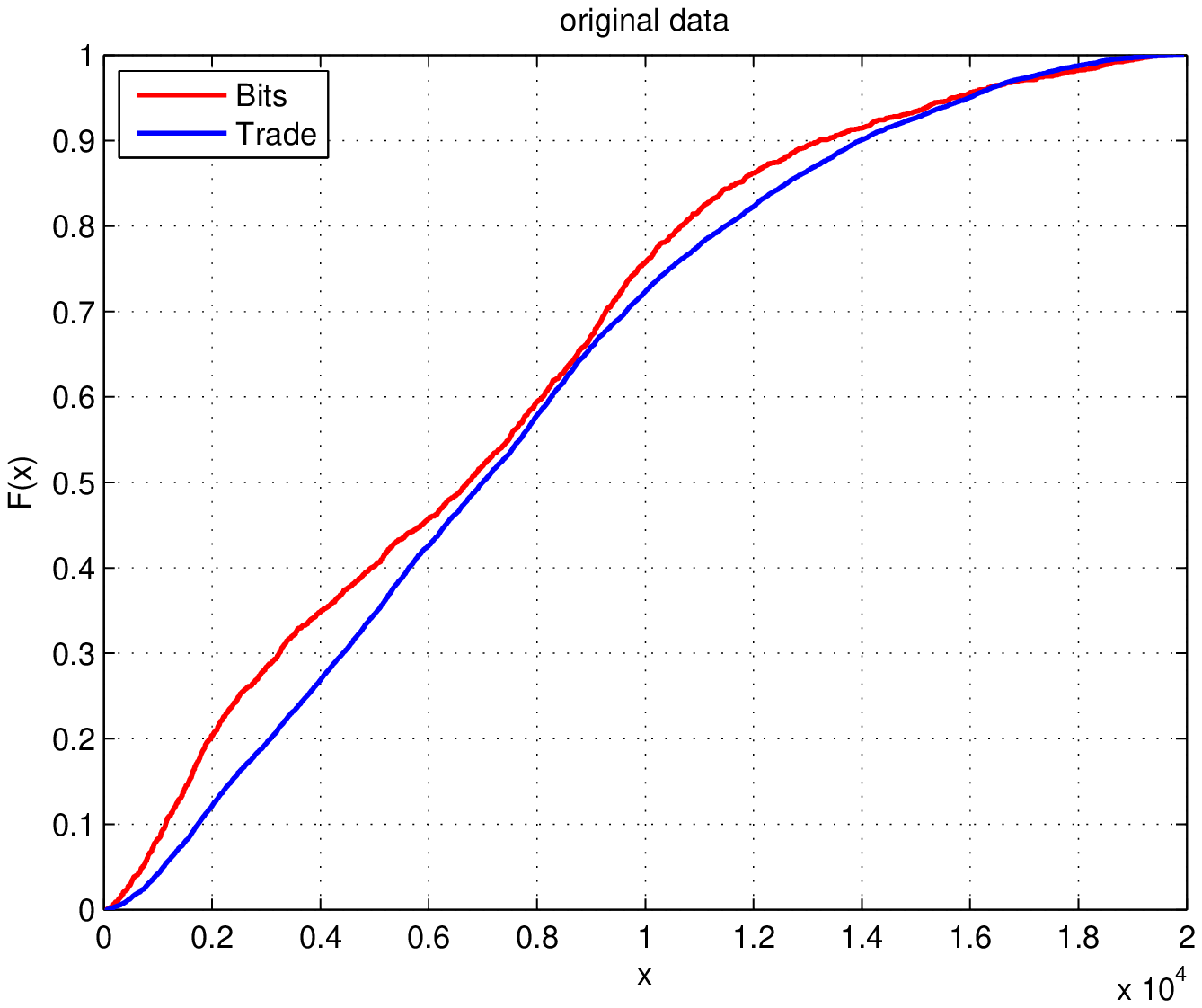} & &
\includegraphics[width=1.5in,height=1.5in,angle=0]{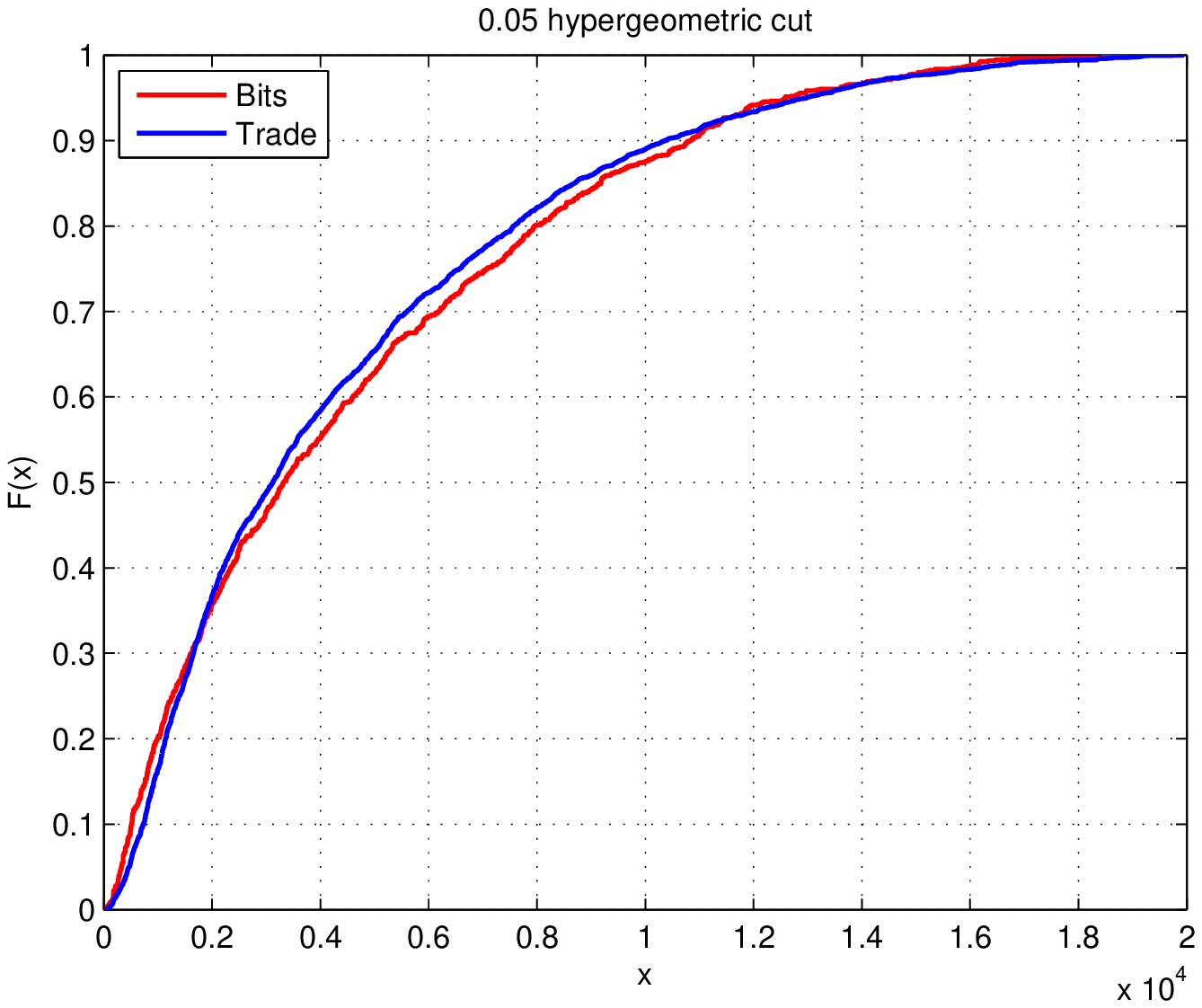} & &
\includegraphics[width=1.5in,height=1.5in,angle=0]{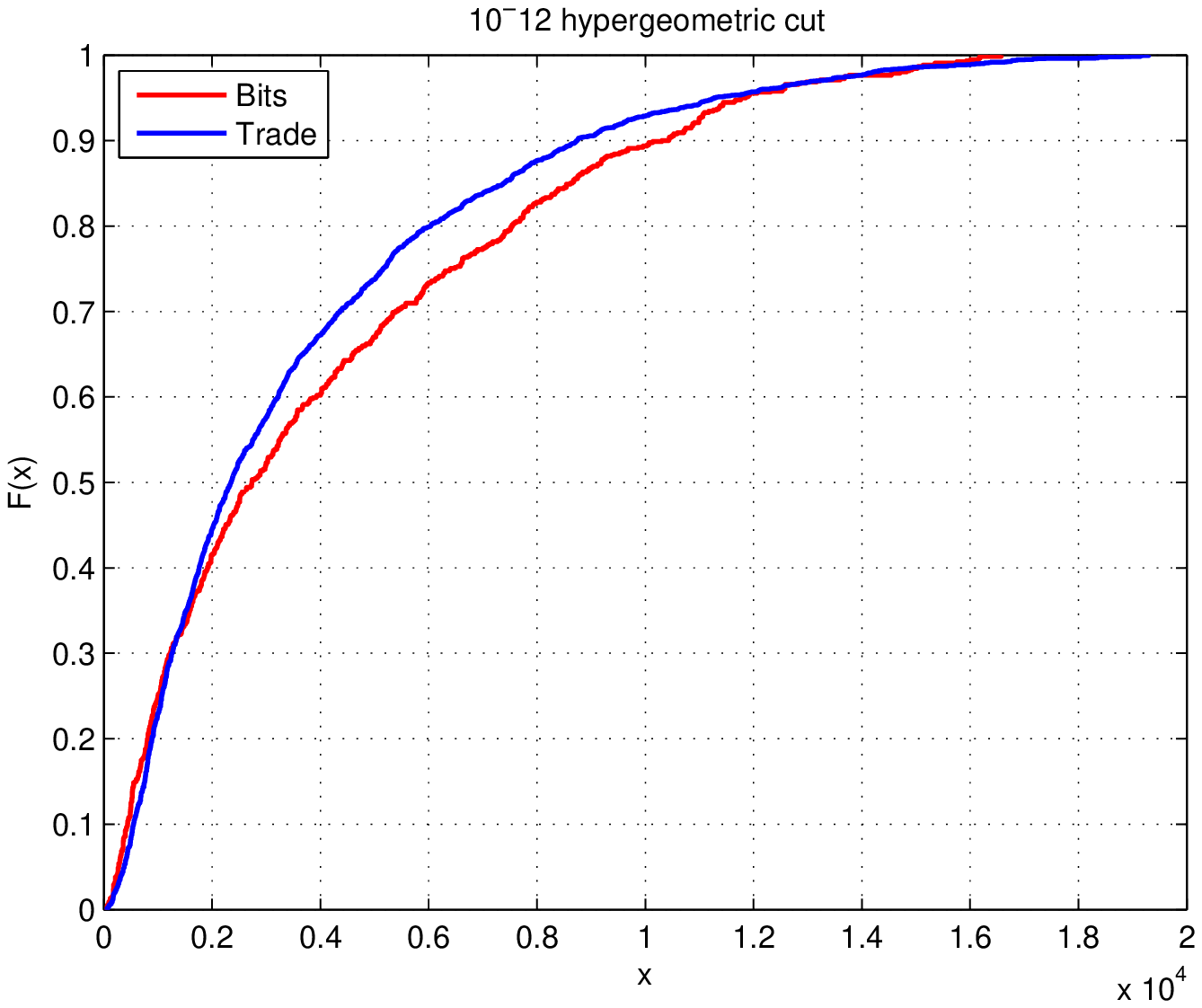} \\
\end{tabular}
\caption{Empirical cumulative distribution for distance between connected pairs. Top row: probability distribution functions; bottom row: cumulative distribution functions. Left panel: original data; center panel: hypergeometric filtering with $P<0.05$; right panel: hypergeometric filtering with $P<10^{-12}$.}
\label{fig:ks}
\end{figure}

As a further step, instead of looking at average distance only, we take a broader view and compare the distribution of distances across the various networks by means of a Kolmogorov-Smirnov test. Once
again, we find evidence of an inversion in the distance ranking between Bits and Trade when moving from the original data to those obtained after hypergeometric filtering. In the former case, we
cannot reject the null hypothesis of the distribution of distances for Bits smaller than that for Trade, whereas the opposite holds true for the  $10^{-12}$ threshold. Figure \ref{fig:ks} shows these
findings by displaying a comparison of the distribution of distances in the two networks.

%
%\begin{figure}[htbf]
%\centering
%\includegraphics[width=5in]{mappa.eps} \\
%\caption{Global computer and trade networks, statistically significant linkages, maximum cut.}
%\label{fig:map}
%\end{figure}

%\paragraph{QAP}
Lastly, another way of testing for the correlation between network structure and distance across countries is by a quadratic assignment procedure  \citep[QAP, see for instance][]{krac87,huar89}. This
entails computing the correlation between the matrix of hypergeometric cumulative probabilities P and the matrix of physical distances across countries, and comparing it with the values obtained once
the rows and columns of P are randomly shuffled.

This correlation turns out to be positive and significant for all four networks. This means that, as distance grows, so does the cumulative hypergeometric probability associated with the link
intensity between the two countries. In other words, a positive correlation implies that more distant countries are connected by links whose intensity is increasingly similar to what our null model
would predict. Another way of looking at the same result is to say that links which are significantly stronger than the hypergeometric benchmark are associated with lower distances. This correlation
is 0.153 for the Bits network but 0.351 for Trade.\footnote{Results do not differ substantially for trade in high-skill (0.360) or low-skill goods (0.349).} Hence, the role played by distance in
hindering strong connections is more evident in the case of trade in physical goods than trade in digital services.

\section{Conclusions}

Analysis of topological properties shows that the Bits network is sparser, less dense, with fewer bilateral links with respect to Trade networks, with differences which are higher in comparison with
low-skill than high-skill product trade. This indicates that the Internet is a polarized network, with large fluxes occurring among a smaller set of countries. All networks share a substantial
disassortative structure but, after hypergeometric filtering, most of the core-periphery links fade out in the Bits network and the underlying structure becomes significantly less disassortative
compared with Trade. Clustering and rich club analysis both confirm this evidence: after filtering the data, we find that many triangles observed in Bits disappear, whereas rich club analysis suggests
that stronger links occur mainly between the core and the periphery in Trade, a larger number of them occurring among hubs in the case of Bits. Lastly, distance plays a much stronger role in shaping
the structure of trade in physical goods than in the case of Bits. When moving from the complete network to an examination only of the strongest links, we do observe that the average distance between
partners increases in the Bits network more than in that of Trade.

This analysis leaves a number of interesting avenues for further research, related to open questions. 
On the methodological ground, one concerns the filtering method based on hypergeometric distribution, and has to do with a way of discriminating better among ``large'' flows. 
The analysis also shows that fluxes tend to cluster around zero, so that beyond a certain threshold it becomes increasingly difficult to distinguish among them. We conjecture that, by manipulating the three distribution
parameters, it should be possible to increase the power of the hypergeometric test. 

A second open question also relates to the use of ``small'' fluxes (i.e., links whose values are significantly smaller than those predicted by a hypergeometric benchmark) to extract further information about  the topological properties of the networks, as well as on their economic and social implications.

On the relationship between trade in goods and digital services we presented some preliminary results on the digital divide and the tyranny of distance that it would be worthwhile to investigate further.
Table \ref{tab:countrieslist} in the appendix shows that all countries with no statistically significant link in the high-skill good network have zero or just one linkage in the digital network.\footnote{A notable exception is Iraq, but the presence of international armed forces, with the associated need of connections with home countries, easily explains the peculiar behavior of that country.} 
This evidence suggests that the development of trade in high-skill goods and digital services are intertwined and mutually reinforce each other.
Moreover, our findings in this paper reveal that long-distance trade can be partially replaced by digital relationships as these lower the costs of monitoring subsidiaries or foreign affiliates.
This can help to explain why ---contrary to expectations--- the digital revolution did not imply the death of distance for trade. 

\bibliographystyle{agsm} 
\bibliography{bits2}

\newpage
\appendix
%\appendixpage
\numberwithin{equation}{section}
\numberwithin{table}{section}
\numberwithin{figure}{section}

\section{Country List}
\label{sec:countries}

\begin{table}[bh!]
\begin{center}
\begin{sideways}
%\begin{sidewaystable}[]
\scriptsize
    \begin{tabular}{llllll}
    \hline
    Afghanistan & Cayman isl.** & Germany & Lebanon & Norway & Suriname* \\
    Albania* & Central African Rep* & Ghana & Liberia* & Oman*  & Sweden \\
    Algeria** & Chile & Gibraltar* & Libya** & Pakistan & Switzerland \\
    Andorra* & China & Greece & Lithuania & Palau* & Syria* \\
    Angola & Cocos isl.** & Greenland** & Macau* & Panama & Taiwan \\
    Anguilla* & Colombia & Grenada** & Macedonia & Papua New Guinea* & Tajikistan* \\
    Antigua and Barbuda & Congo* & Guatemala & Madagascar* & Paraguay* & Tanzania \\
    Argentina & Congo, Dem. Rep.** & Guinea* & Malawi* & Peru  & Thailand \\
    Armenia** & Costa rica** & Guyana* & Malaysia & Philippines & Togo* \\
    Aruba* & Cote d'ivoire** & Haiti* & Maldives* & Poland & Tonga* \\
    Australia & Croatia & Honduras & Mali*  & Portugal & Trinidad and Tobago \\
    Austria & Cuba*  & Hong kong & Malta** & Qatar* & Tunisia** \\
    Azerbaijan & Cyprus** & Hungary & Marshall isl.* & Romania & Turkey \\
    Bahamas & Czech Rep. & Iceland* & Mauritania* & Russian federation & Turkmenistan* \\
    Bahrain** & Denmark & India & Mauritius** & Rwanda** & Tuvalu* \\
    Bangladesh** & Djibouti* & Indonesia & Mexico & St. kitts and nevis & Uganda** \\
    Barbados & Dominica* & Iran  & Moldova** & St. Lucia & Ukraine \\
    Belarus** & Dominican Rep. & Iraq  & Mongolia* & St. Pierre and Miquelon* & UAE \\
    Belgium-Luxembourg & Ecuador & Ireland & Montserrat** & St. Vincent and Grenadines** & UK \\
    Belize** & Egypt & Israel & Morocco & Samoa* & USA \\
    Bermuda & El salvador & Italy & Mozambique & Sao Tome and Principe* & Uruguay \\
    Bhutan* & Equatorial guinea** & Jamaica & Myanmar** & Saudi arabia & Uzbekistan** \\
    Bolivia & Estonia & Japan & Nepal* & Senegal & Vanuatu* \\
    Bosnia Herzegowina** & Ethiopia* & Jordan** & Netherlands & Seychelles** & Venezuela \\
    Brazil & Falkland isl.* & Kazakhstan & Net. Antilles & Singapore & Vietnam \\
    Brunei darussalam* & Fiji*  & Kenya & New Caledonia* & Slovakia & Virgin isl. (British)** \\
    Bulgaria & Finland & Kiribati* & New Zealand & Slovenia & Yemen* \\
    Burkina faso* & France & Korea, Rep. & Nicaragua** & Solomon isl.* & Zambia* \\
    Burundi* & French polynesia & Kuwait** & Niger* & South African CU & Zimbabwe** \\
    Cambodia** & Gabon* & Kyrgyzstan* & Nigeria & Spain &  \\
    Cameroon* & Gambia* & Laos*  & Norfolk isl.* & Sri lanka** &  \\
    Canada & Georgia** & Latvia & North. Mariana isl.** & Sudan** &  \\
    \hline
\end{tabular}
 \label{tab:countrieslist}
\end{sideways}
\end{center}
  \caption{\scriptsize * Countries without statistically significant outgoing links in the computer network; ** Country with only one statistically significant outgoing link in the computer network. All countries without statistically significant
export links in the skill-intensive trade network have less that two outgoing links in the computer network, except Iraq
(2 links).}
  
%\end{sidewaystable}
\end{table}

\newpage

%\section{Global computer and trade network map}
%\label{sec:map}
%
%\begin{figure}[htbf]
%\centering
%\includegraphics[width=8in,angle=90]{mapP.eps} \\
%\caption{Global computer and trade networks, statistically significant linkages, cut $10^{-12}$.}
%\label{fig:map2}
%\end{figure}

\end{document}